\documentclass[pre,final,twocolumn,showpacs,showkeys]{revtex4-2}
\usepackage{amsmath}
\usepackage{graphicx}
\usepackage{amsfonts}
\usepackage{amssymb}
\usepackage{physics}
\usepackage{subfigure}

\def\identity{\leavevmode\hbox{\small1\kern-3.2pt\normalsize1}}

\begin{document}

\title{Optimal Conditions for Environment-Assisted Quantum Transport on the Fully Connected Network }

\author{Sam Alterman}
\altaffiliation[Present address: ]{Department of Physics and Astronomy, Tufts University, Medford, MA 02155}
\author{Justin Berman}
\altaffiliation[Present address: ]{Department of Physics, University of Michigan, Ann Arbor, MI 48109-1040 }
\author{Frederick W. Strauch} \email[Electronic address: ]{Frederick.W.Strauch@williams.edu}
\affiliation{Department of Physics \\ Williams College, Williamstown, MA 01267}

\date{\today}

\begin{abstract}
We present a theoretical analysis of the efficiency and rate of excitation transport on a network described by a complete graph in which every site is connected to every other.  The long-time transport properties are analytically calculated for networks of arbitrary size that are symmetric except for the trapping site, start with a range of initial states, and are subject to dephasing and excitation decay.  Conditions for which dephasing increases transport are identified, and optimal conditions are found for various physical parameters.  The optimal conditions demonstrate robustness and a convergence of timescales previously observed in the context of light-harvesting complexes.  
\end{abstract} 
\maketitle

\newcommand{\subs}[1]
{ 
	\mbox{\scriptsize{#1}}
}

\section{Introduction}

The role of quantum coherence in photosynthetic systems is a topic of long-standing \cite{fleming1994primary} and continued \cite{scholes2011lessons, *fassioli2014photosynthetic, *levi2015quantum,*wang2019quantum} interest, and forms one of the pillars of the new field of quantum biology \cite{ball2011physics, *huelga2013vibrations, *marais2018future}.  Much of this interest stems from the experiment of Engel {\em et al.}~\cite{engel2007evidence}, in which long-lived time-resolved oscillations were observed using ultrafast nonlinear spectroscopy.  These oscillations were interpreted as evidence for a coherent wavelike transfer of energy whose speed and efficiency could be understood as something akin to Grover's quantum search algorithm \cite{grover1997quantum}.   

The specific light-harvesting complex studied in \cite{engel2007evidence}, the Fenna-Mathews-Olsen (FMO) complex, has been the subject of intense investigation \cite{jang2018delocalized}.  This body of work shows that a simple wavelike interpretation of the excitation energy transfer (EET) is not correct.  Early theoretical investigations quickly showed that the dynamics of EET in FMO is in a regime where environmental noise is both significant and advantageous \cite{mohseni2008environment}.  This was something of a surprise, as such noise usually causes a loss of coherence through dephasing.  Subsequent work confirmed that the role of coherence in FMO was limited and the transport could not be interpreted as quantum search \cite{rebentrost2009role,hoyer2010limits}.    Indeed, many now argue that quantum coherence is not essential to photosynthesis \cite{wilkins2015quantum, *duan2017nature, *cao2020quantum, *zerah2021photosynthetic}.  Nevertheless, the general phenomenon where noise assists transport \cite{plenio2008dephasing}, also known as environment-assisted quantum transport (ENAQT) \cite{rebentrost2009environment}, is a topic of great interest.    

The discovery of ENAQT encouraged a more expansive study of EET in networks beyond FMO, with an eye towards the design of engineered networks.   Rebentrost {\em et al.} \cite{rebentrost2009environment} studied EET on a binary tree in which each site has a random energy, known as on-site disorder.  For sufficiently strong disorder, the eigenstates of the system will be subject to Anderson localization, which suppresses transport.  However, given sufficient dephasing, transport can be recovered.  It is interesting to note that an example of this phenomenon was seen in the conductivity of electron transport on a binary tree in work by Jonson and Girvin in 1979 \cite{jonson1979electron, *girvin1980dynamical}.   The existence of ENAQT for finite ordered chains was demonstrated in \cite{kassal2012environment}, while the conditions for optimal EET on the binary tree and the hypercube were analyzed in \cite{novo2016disorder}.  ENAQT has since been experimentally observed in trapped-ion \cite{maier2019environment}, superconducting \cite{potovcnik2018studying}, and optical cavity \cite{viciani2015observation} systems.

One of the characteristic features of ENAQT, in networks or FMO, is that optimal transport involves a convergence of relevant timescales \cite{mohseni2014energy}.  This convergence, which yields an inherent robustness \cite{shabani2014numerical}, has been dubbed the ``Goldilocks effect'' \cite{lloyd2011quantum}.  The possibility that such a dynamical regime is important for photosynthesis was suggested some time ago \cite{skourtis1992new}. 

Most studies of ENAQT for large networks have involved numerical work.  However, analytical results are known for systems with a few sites \cite{plenio2008dephasing, cao2009optimization, mulken2010environment, kassal2012environment}.  One particularly nice network is the complete graph, in which every site is connected to every other site; this is is also known as the fully connected network (FCN).  Transport on this network admits exact treatment, for arbitrarily large size, due to its high degree of symmetry.  Using this fact, Caruso {\em et al.} \cite{caruso2009highly} studied EET on this network and found evidence for ENAQT in certain limiting regimes. 

\begin{figure}
\includegraphics[width=3 in]{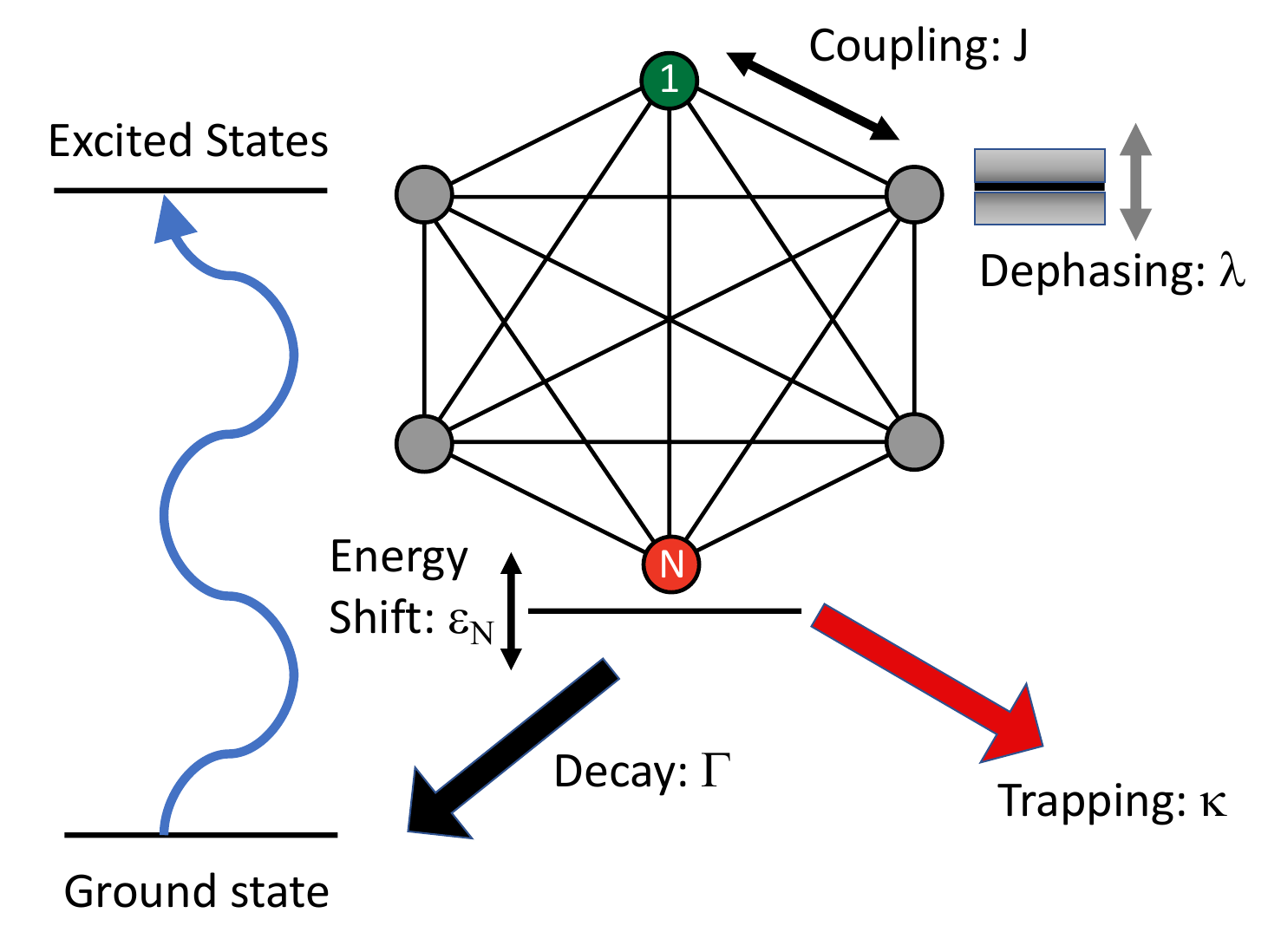} 
\label{fig1}
\caption{Model for excitation energy transfer on the fully connected network.  An initial excitation is generated (blue arrow) from a common ground state to a set of excited states (filled circles), such as state $|1\rangle$ (in green).  Each state is coupled to every state with a coupling strength $J$ and decays back to the ground state with rate $\Gamma$.  Energy is collected via state $|N\rangle$ (in red), which decays to a trapping site with rate $\kappa$.  This state has an energy shift $\varepsilon_{N}$ with respect to the other excited states.  Finally, all states are subject to a dephasing process with characteristic rate $\lambda$.  }
\end{figure}

In this paper we extend their model to enable a broader analysis of ENAQT on the FCN.  We find analytical results for both the efficiency and the rate of EET for the entire range of physical parameters.   By including an energy shift, our extended model contains the continuous-time version \cite{Farhi1998analog} of Grover's quantum search algorithm.  By varying the initial state, we are able to identify the role of initial state coherence on the transfer.   We find the optimal regimes for ENAQT on the FCN and demonstrate an analytical convergence of timescales.  Our findings show that ENAQT and Grover search occur in distinct dynamical regimes.  These results serve to elucidate the nature of optimal and robust transport in a rich dynamical system.   

This paper is organized as follows.  In Section II we review the dynamical framework we use for studying EET on the FCN.  The main results for ENAQT are presented in Section III, in which the efficiency and rate of transfer are found for networks of arbitrary size.  This analysis uses a reduced set of equations for the density matrix, which are derived in the Appendix.  The role of initial state coherence is explored in Section IV, and the Grover search limit is identified.   The optimal conditions for ENAQT on the FCN are derived in Section V.   We conclude in Section VI. 

\section{Theoretical Model of EET}

Our model for EET follows that of \cite{rebentrost2009environment, caruso2009highly}.  The key features of this model are illustrated in Fig.~1, and can be summarized as follows.  An excitation is produced in a network of sites.  This excitation can move throughout the network, and will ultimately decay (back to some common ground state) or be trapped at a reaction site.  The network Hamiltonian is given by 
\begin{equation}
H_0=\sum_{j=1}^N \varepsilon_j \ket{j}\bra{j} + \sum_{j,k=1;\, j\neq k}^N V_{jk}\ket{j}\bra{k}, \label{eq:FCNDiracham}
\end{equation}
where $\ket{j}$ represents an excitation in the $j^\text{th}$ site, $\varepsilon_j$ are the site energies and $V_{jk}$ are the couplings between sites.   For the fully connected network, we set $V_{jk} = J$, and we further restrict the site energies $\varepsilon_j = 0$ for all sites except $j = N$, which we take as the trapping site.  Trapping at site $N$ is modeled by the damping term
\begin{equation}
H_\text{trap}=-i \ \kappa \left \vert N\left\rangle \right \langle N\right\vert , 
\end{equation}
while the decay of each excitation is given by 
\begin{equation}
H_\text{decay}=-i \ \Gamma \sum_{j=1}^N \left \vert j\left\rangle \right \langle j\right\vert. 
\end{equation}

In addition, each site is affected by noise from its local environment.  We model this by a purely Markovian dephasing process, so that the density matrix evolves via the Lindblad equation
\begin{align}
\frac{d \rho}{dt} = & -i (H \rho - \rho H^{\dagger}) \nonumber \\
& + \lambda \sum_j \left(L_j\rho L_j^\dagger-\frac{1}{2}L_j^\dagger L_j\rho-\frac{1}{2}\rho L_j^\dagger L_j\right),
\label{dmateq}
\end{align}
where the total Hamiltonian is
\begin{equation}
H = H_0 + H_\text{trap} + H_\text{decay},
\end{equation}
 the Lindblad operators are $L_j = \ket{j} \bra{j}$, and $\lambda$ is the dephasing rate, taken to be uniform for all sites.  (Note that we are using units such that $\hbar = 1$).
 
The efficiency of EET is given by the total population trapped at site $N$
\begin{equation}
\eta = 2 \kappa \int_{0}^{\infty} \rho_{NN}(t) \ dt,
\end{equation}
while the mean transfer time is defined by
\begin{equation}
\tau = \frac{2}{\eta} \kappa \int_{0}^{\infty} t \ \rho_{NN}(t) \ dt.
\end{equation}
By writing the density matrix equation in superoperator form 
\begin{equation}
\frac{d \rho}{d t} = \mathcal{L} \rho, \label{dmatsuper}
\end{equation}
the efficiency can be evaluated \cite{rebentrost2009role} as
\begin{align}
\eta &= 2 \kappa \int_{0}^{\infty} \bra{N} e^{\mathcal{L} t} \rho(0) \ket{N} dt \nonumber \\
&= 2 \kappa \bra{N} \left[ \mathcal{L}^{-1} e^{\mathcal{L} t} \rho(0) \right]_{0}^{\infty} \ket{N} \nonumber \\
&= - 2 \kappa \bra{N} \mathcal{L}^{-1} \rho(0) \ket{N}
\label{efficiency}
\end{align}
where we have used the fact that matrix elements of $e^{\mathcal{L} t}$ go to zero as $t \to \infty$.  The mean transfer time can similarly be evaluated as
\begin{align}
\tau &= \frac{2}{\eta} \kappa \int_{0}^{\infty} \bra{N} t e^{\mathcal{L} t} \rho(0) \ket{N} dt \nonumber \\
&= \frac{2}{\eta} \kappa \bra{N} \left[ \mathcal{L}^{-2} (\mathcal{L} t - I)  e^{\mathcal{L} t} \rho(0) \right]_{0}^{\infty} \ket{N} \nonumber \\
&=  \frac{2 \kappa}{\eta} \bra{N} \mathcal{L}^{-2} \rho(0) \ket{N}.
\label{transfert}
\end{align} 
While $\eta$ and $\tau$ are the usual parameters studied in EET, their combination $\mathcal{R} = \eta / \tau$, the overall rate of energy transfer, is also of interest.  

\section{Solution for the FCN}

Given this theoretical model for EET, we solve the following problem:  Starting an initial condition $\rho(0)$ on the FCN, find $\eta$ and $\tau$ in terms of the physical parameters $J$ (the coupling between sites), $\varepsilon_N$ (the energy of the trapping site), $\kappa$ (the trapping rate), $\Gamma$ (the excitation decay rate), and $\lambda$ (the dephasing rate).  

We use the dimensional reduction method introduced by Caruso {\it et al.} \cite{caruso2009highly} to study networks with arbitrary size $N$.  In their work, certain limiting results for $\eta$ were found using the Laplace transform of the reduced equations of motion.  Here we use the analytical forms of both $\eta$ in Eq.~(\ref{efficiency}) and $\tau$ in Eq.~(\ref{transfert}) and matrix manipulations to find the general solution.  We further extend the model to include an energy shift $\varepsilon_N$ for the trapping site, and consider a range of initial conditions for $\rho(0)$.  

The reduced equations are derived in the Appendix, and can be written in matrix form 
\begin{equation}
\frac{d {\bf v}}{dt}=\mathcal{L} {\bf v}
\end{equation}
where
\begin{widetext}
\begin{equation}
{\bf v} = \left(\begin{array}{c}
\rho_{NN}\\
X\\
Y\\
S\\
T\end{array}\right), \ \
\mathcal{L}=\left(\begin{array}{c c c c c}
-2(\Gamma+\kappa) & 0 & 2J & 0 & 0\\
-(\kappa-\lambda) & -(\lambda+2\Gamma+\kappa) & (JN-\varepsilon_N) & 0 & 0\\
-\varepsilon_N & -(JN-\varepsilon_N) & -(\lambda+2\Gamma+\kappa) & J & 0\\
0 & -2\kappa & -2\varepsilon_N & - (\lambda+2\Gamma) & \lambda\\
-2\kappa & 0 & 0 & 0 & -2\Gamma \end{array}\right),
\label{matrixdefs}
\end{equation}
\end{widetext}
and $X$, $Y$, $S$, and $T$ are linear combinations of the elements of the density matrix (with expressions found in the Appendix).  

\subsection{Transport Properties}

Using the reduced equations of motion, the efficiency is
\begin{equation}
\eta = - 2 \kappa \left[ \mathcal{L}^{-1} {\bf v}(0) \right]_1,
\end{equation}
and the mean transfer time is
\begin{equation}
\tau = \frac{2 \kappa}{\eta} \left[ \mathcal{L}^{-2} {\bf v}(0) \right]_1,
\end{equation}
where $\mathcal{L}$ is defined in Eq.~(\ref{matrixdefs}),  ${\bf v}(0)$ is found from the initial conditions of the density matrix, and the subscript means that we are evaluating the first element of the vector (corresponding to $\rho_{NN}$).  The required matrix manipulations are performed in Mathematica.  We consider an initial pure state of the form
\begin{equation}
|\psi(t=0)\rangle = \frac{1}{\sqrt{S}} \sum_{j=1}^S |j\rangle,
\end{equation} 
where $S < N$ is the number of terms in superposition.  This state allows for some initial coherence, but no population in or coherence with the trapping site $|N\rangle$.  The reduced vector for this initial state is
\begin{equation}
{\bf v}(0) =  \left(\begin{array}{c}
0\\
0\\
0\\
S\\
1\end{array}\right)\label{eq:vdef}.
\end{equation}
Note that $S = 1$ would also describe any localized initial state $|j\rangle$, $j \ne N$, or a statistical mixture of such states.   

The resulting efficiency can be written in the following form:
\begin{equation}
\eta = \frac{1}{\alpha_1 + \alpha_2  (\Delta^2 / J^2)},
\label{effeq}
\end{equation}
where
\begin{align}
\alpha_1 =& \frac{ \lambda (\kappa + N \Gamma) + 2 (N-1) \Gamma (\kappa + 2 \Gamma)}{\kappa (\lambda + 2 S \Gamma)}  \nonumber \\
& + \frac{\Gamma (\kappa + \Gamma) (\lambda + 2 \Gamma) (\kappa + \lambda + 2 \Gamma)}{J^2 \kappa (\lambda + 2 S \Gamma)},
\label{alpha1general}
\end{align}
\begin{equation}
\alpha_2 = \frac{\Gamma (\kappa + \Gamma) (\lambda + 2 \Gamma) }{\kappa (\kappa + \lambda + 2 \Gamma)(\lambda + 2 S \Gamma)},
\label{alpha2general}
\end{equation}
and we have defined the detuning
\begin{equation}
\Delta = \varepsilon_N - J (N-2).
\end{equation}
We will analyze $\eta$ further in Sec. III.C.

The transfer time can also be evaluated, but it turns out the transfer rate is more convenient to analyze:
\begin{equation}
\mathcal{R} = \frac{\eta}{\tau} = \frac{2 \kappa}{\beta_1 + \beta_2 (\Delta^2/J^2)},
\end{equation}
where
\begin{align}
\beta_1 =& \frac{2 (N-S-1) \kappa \lambda + N \lambda^2 + 8 (N-1) \Gamma (\lambda + S \Gamma)}{(\lambda+ 2 S \Gamma)^2} \nonumber \\
& + \ \frac{ \Gamma \left[ (\kappa + \Gamma) (2 \kappa + 4 \lambda + 8 \Gamma) + (\lambda + 2 \Gamma) (\kappa + \lambda + 2\Gamma)\right]}{J^2 (\lambda + 2 S \Gamma)} \nonumber \\
 & + \ \frac{\lambda (\lambda + 2 \Gamma)(\kappa +\Gamma) (\kappa + \lambda + 2 \Gamma)}{ J^2 (\lambda + 2 S \Gamma)^2}
 \label{beta1general}
\end{align}
and
\begin{align}
\beta_2 =& \frac{ \lambda (\lambda + 2 \Gamma) (\kappa + \Gamma)}{(\kappa + \lambda + 2 \Gamma) (\lambda + 2 S \Gamma)^2} \nonumber \\
& + \ \frac{2 \Gamma \kappa (\kappa + \Gamma)}{(\kappa + \lambda + 2 \Gamma)^2 (\lambda +2 S \Gamma)} \nonumber \\
& + \ \frac{\Gamma (\lambda + 2\Gamma)}{(\kappa + \lambda + 2 \Gamma)(\lambda + 2 S \Gamma)}.
\label{beta2general}
\end{align}
We will analyze $\mathcal{R}$ further in Sec. III.D.

These expressions constitute the main results of this paper.  These exact solutions for $\eta$ and $\mathcal{R}$ are somewhat difficult to analyze, for general values of $\Delta$, $\kappa$, $\lambda$, $\Gamma$ (we treat $J$, $N$, and $S$ as fixed).  However, these results extend and generalize previous analytical results \cite{cao2009optimization,caruso2009highly} and reduce to them in various limits.  The inclusion of $S$ and $\Delta$ allows for an additional limit, namely the continuous-time version \cite{Farhi1998analog} of Grover's quantum search algorithm.  We will first look at the general features of $\eta$ and then consider the analytical limits in remainder of this section.

\begin{figure*}
\includegraphics[width=6 in]{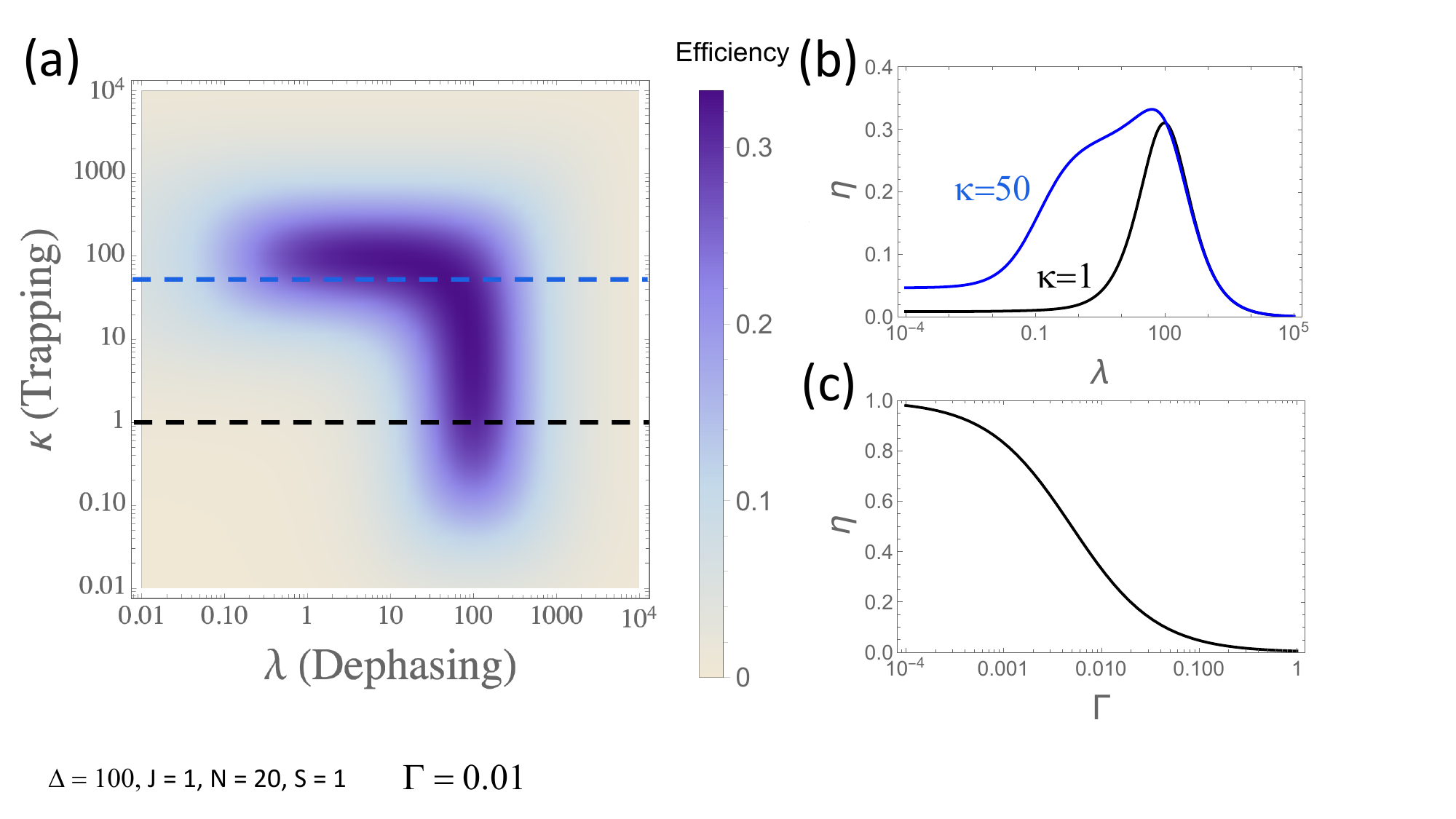}
\label{fig2}
\caption{(a) Efficiency $\eta$ as a function of $\lambda$ and $\kappa$ for transport on a fully connected network with $N=20$, $S=1$, $\Gamma = 0.01$, $J=1$, and $\Delta = 100$.  The dashed lines indicate the slices $\kappa = 1$ (lower, black) and $\kappa = 50$ (upper, blue).  The global optimum $\eta = 0.33$ occurs for $\lambda = 56$ and $\kappa = 44$. (b) The efficiency as a function of $\lambda$ for $\kappa = 1$ (lower, black) and $\kappa = 50$  (upper, blue). (c) The efficiency as a function of $\Gamma$ for $\kappa = \lambda = 50$.}
\end{figure*}

\begin{figure*}
\includegraphics[width=6 in]{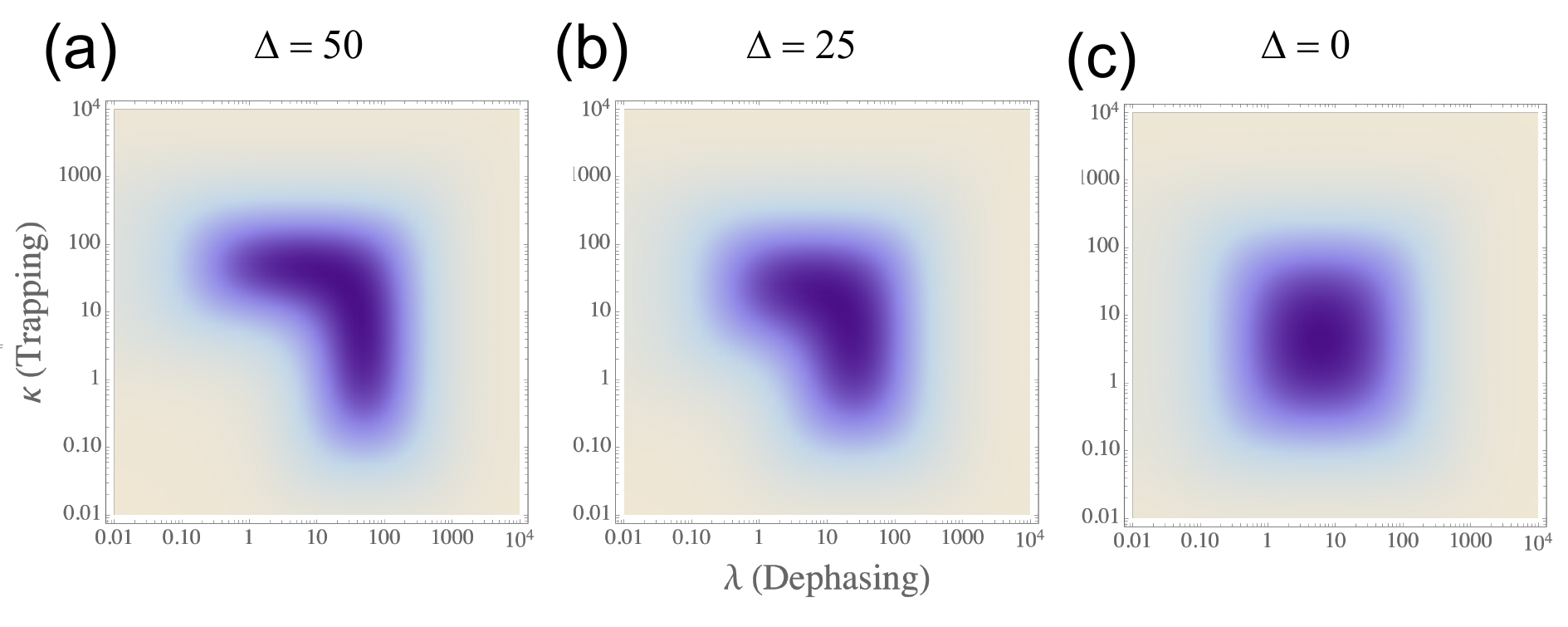}
\label{fig3}
\caption{Efficiency $\eta$ as a function of $\lambda$ and $\kappa$ for transport on a fully connected network with $N=20$, $S=1$, $\Gamma = 0.01$, $J=1$ for (a) $\Delta = 50$, with optimum $\eta = 0.49$ at $\lambda = 20$ and $\kappa = 22$, (b) $\Delta = 25$, with optimum $\eta = 0.65$ at $\lambda = 15.4$ and $\kappa = 11.7$ and (c) $\Delta = 0$, with optimum $\eta = 0.83$ at $\lambda = 6$ and $\kappa = 4.5$.}
\end{figure*}

\subsection{Typical Efficiency for ENAQT}
To proceed, we note that previous work suggests that the regime for ENAQT is in the regime $\Gamma < J < \Delta$, to which we direct our attention.  Here we consider a typical case with specific parameters of $\Gamma = 0.01$, $J=1$; we will derive analytical expressions in the following sections.  

The efficiency for a network of size $N=20$, initially localized at a single site (with $S=1$), is shown as a function of $\lambda$ and $\kappa$  in Fig.~2(a) (with $\Delta = 100$).  This exhibits the characteristic shape of ENAQT \cite{rebentrost2009environment}, with an enhancement of transport with dephasing.   This is shown in more detail in Fig.~2(b), where slices of the efficiency are shown as functions of $\lambda$ for $\kappa = 1$ and $\kappa = 50$.  Each shows an enhancement with dephasing with a maximum efficiency for $\lambda \approx \Delta$.  The optimal enhancement occurs near $\kappa \approx \lambda \approx \Delta/2$, but persists over a broad range of parameters.  The efficiency decreases for increasing $\Gamma$, as shown in Fig.~2(c).  

We also observe that ENAQT persists for decreasing $\Delta$, as shown in Fig.~3.  The location of optimal enhancement decreases towards $\kappa \approx \lambda \approx \sqrt{N} J$, while the efficiency increases.  In fact, Eq.~(\ref{effeq}) shows that the efficiency is always maximized by setting $\Delta = 0$, and thus $\varepsilon_N = J (N-2)$.  

\subsection{Analytical Results for the Efficiency}

Many analytical limits can be obtained from the general solution.  We begin by comparing our results to those of Caruso {\em et al.}~\footnote{Our result should agree with their Eq.~(A28) for a correct calculation of their denominator.}.   First, we consider the case of no dephasing, or $\lambda = 0$.  In this case Eq.~(\ref{alpha1general}) reduces to
\begin{equation}
\alpha_1 = \frac{N-1}{S} (1 + 2 \Gamma/\kappa) + \frac{\kappa \Gamma}{S J^2} (1+ \Gamma/\kappa) (1 + 2\Gamma/\kappa)
\end{equation} 
and Eq.~(\ref{alpha2general}) to 
\begin{equation}
\alpha_2 =  \frac{\Gamma}{S \kappa} (1 + \Gamma/\kappa) (1 + 2\Gamma/\kappa)^{-1}.
\end{equation}
The resulting efficiency agrees with Eq.~(10) of \cite{caruso2009highly} for $S=1$ and $\Delta = -J(N-2)$ (the model used there has $\varepsilon_N = 0$).  We also see, by setting $\Gamma = 0$, that
\begin{equation}
\alpha_1 \to \frac{N-1}{S} \ \ \mbox{and} \ \ \alpha_2 \to 0,
\end{equation} 
and thus
\begin{equation}
\eta \to \frac{S}{N-1} \ \ \mbox{for} \ \ \lambda = 0, \Gamma \to 0.
\label{effcoherence}
\end{equation}
For $S=1$, this agrees with Eq.~(5) of \cite{caruso2009highly}.  

However, when the coherence of the initial state increases, the efficiency increases, reaching unity for $S=N-1$.  The fully coherent limit $\lambda = 0, \Gamma \to 0, \kappa \to 0$ of our theoretical model corresponds to the Hamiltonian dynamics of a continuous-time quantum walk on the complete graph  \cite{childs2004spatial}.  This quantum walk, in turn, corresponds to the continuous-time version of Grover's search algorithm \cite{Farhi1998analog}, with the trapping site corresponding to the desired target state.  The probability of success is maximized and equals unity for $S = N-1$, $\Delta = 0$, and at the measurement time $T_0 = \pi / (2 J \sqrt{N-1})$.  We will discuss this further in Sec. IV.

Second, we consider the case of no excitation decay, or $\Gamma = 0$.  In this case Eqs.~(\ref{alpha1general}) and (\ref{alpha2general}) reduce to
\begin{equation}
\alpha_1= 1 \ \ \mbox{and} \ \ \alpha_2 = 0,
\end{equation}
so that
\begin{equation}
\eta = 1 \ \ \mbox{for} \ \ \Gamma=0.
\end{equation}
This efficiency agrees with Eq.~(A30) of \cite{caruso2009highly}.  However, we observe that the order of the limits $\Gamma \to 0$ and $\lambda \to 0$ matters.  We will return to this important point in Sec. III. D.

We can also compare our results to those of Cao and Sibley \cite{cao2009optimization}, who present exact results for $N=2$ and $S=1$. In this case Eq.~(\ref{alpha1general}) reduces to
\begin{equation}
\alpha_1 = \frac{\kappa + 2 \Gamma}{\kappa} +  \frac{\Gamma (\kappa + \Gamma)( \kappa + \lambda + 2 \Gamma)}{J^2 \kappa}
\end{equation}
and Eq.~(\ref{alpha2general}) to
\begin{equation}
\alpha_1 = \frac{\Gamma (\kappa + \Gamma)}{\kappa (\kappa + \lambda + 2\Gamma)}.
\end{equation}
The resulting efficiency agrees with their Eq.~(5).  We will consider their results for $N=3$ in the following section.

\subsection{Analytical Results for the Transfer Rate}

In addition to observing ENAQT in the efficiency, it can be also observed in the transfer rate $\mathcal{R}$.  This is especially relevant when the efficiency is close to unity \cite{cao2009optimization, mulken2010environment}.  Thus, we consider the cases considered above.   

First, for the $\lambda = 0$ case, Eq.~(\ref{beta1general}) reduces to
\begin{equation}
\beta_1 = \frac{2 (N-1)}{S} + \frac{1}{S} \frac{\kappa^2}{J^2} \left[1 + 6 (\Gamma/\kappa)+ 6 (\Gamma/\kappa)^2 \right],
\end{equation}
and Eq.~(\ref{beta2general}) to
\begin{equation}
\beta_2 = \frac{1}{S} (1 + 2\Gamma/\kappa)^{-2} \left[ 1+ 2 (\Gamma/\kappa) + 2 (\Gamma/\kappa)^2 \right].
\end{equation}
Taking the limit $\Gamma \to 0$, we obtain the coherent transfer rate
\begin{equation}
\mathcal{R} \to \frac{2 S \kappa}{ 2 (N-1) + (\Delta^2 + \kappa^2)/J^2} \ \ \mbox{for} \ \ \lambda = 0, \Gamma \to 0.
\label{Rcoh}
\end{equation}
The maximal rate, as a function of $\kappa$, occurs for $\kappa = \sqrt{2 (N-1)J^2 + \Delta^2}$, so that
\begin{equation}
\mathcal{R}_{\text{max}} = \frac{S J^2}{\sqrt{2(N-1)J^2 + \Delta^2}} \ \ \mbox{for} \ \ \lambda = 0, \Gamma \to 0.
\label{Rmax}
\end{equation}
In this form, we see that the initial coherence $S$ speeds up the overall transfer, up to a factor of $N-1$.  This speedup corresponds to the coherence enhancement of the efficiency seen in Eq.~(\ref{effcoherence}).  In fact, setting $S = N-1$, $\Delta=0$ and making the replacement $J \to E/N$ \cite{Farhi1998analog} we obtain $\mathcal{R}_{\text{max}} =  (E/N) \sqrt{(N-1)/2} \sim E/\sqrt{N}$, in agreement with the performance of the continuous-time search algorithm. 

Second, for the $\Gamma = 0$ case, Eq.~(\ref{beta1general}) reduces to
\begin{equation}
\beta_1 = 2 (N-S-1) \frac{\kappa}{\lambda} + N  + \frac{\kappa (\kappa + \lambda)}{J^2}
\label{beta1nodecay}
\end{equation}
and Eq.~(\ref{beta2general}) to
\begin{equation}
\beta_2 = \frac{\kappa}{(\kappa + \lambda)}.
\label{beta2nodecay}
\end{equation}
For $N=3$ and $S=1$, this resulting rate agrees with Eq.~(22) of Cao and Sibley \cite{cao2009optimization}.  We will consider the optimal rate in Sec. VI.

We can also establish the approximate relationship between the efficiency and the transfer rate, in the regime of slow decay, used in \cite{cao2009optimization}.  By Taylor expanding the general expressions for $\alpha_1$ and $\alpha_2$ in Eqs.~(\ref{alpha1general}-\ref{alpha2general}) with respect to $\Gamma$, we find
\begin{equation}
\alpha_1 \approx 1 + \frac{\Gamma}{\kappa} \beta_1 
\end{equation}
and
\begin{equation}
\alpha_2 \approx \frac{\Gamma}{\kappa} \beta_2,
\end{equation}
where $\beta_1$ and $\beta_2$ are given by the $\Gamma=0$ results in Eq. (\ref{beta1nodecay}-\ref{beta2nodecay}). 
Thus, we have
\begin{equation}
\eta \approx \frac{1}{1 + 2 \Gamma / \mathcal{R}_0},
\label{etaweakdecay}
\end{equation}
where $\mathcal{R}_0$ is the rate evaluated for $\Gamma=0$.  This form for $\eta$ agrees with the behavior seen in Fig.~2(c) and the optimal values for Figs.~2 and 3. We will use this approximation in Sec. V.  

\section{Coherence Enhancement}

We now consider the role of initial coherence in the density matrix, here parametrized by $S$.  The efficiency as a function of $\lambda$ and $\kappa$ is shown in Fig. 4, for various values of $S$.
\begin{figure*}
\includegraphics[width=6 in]{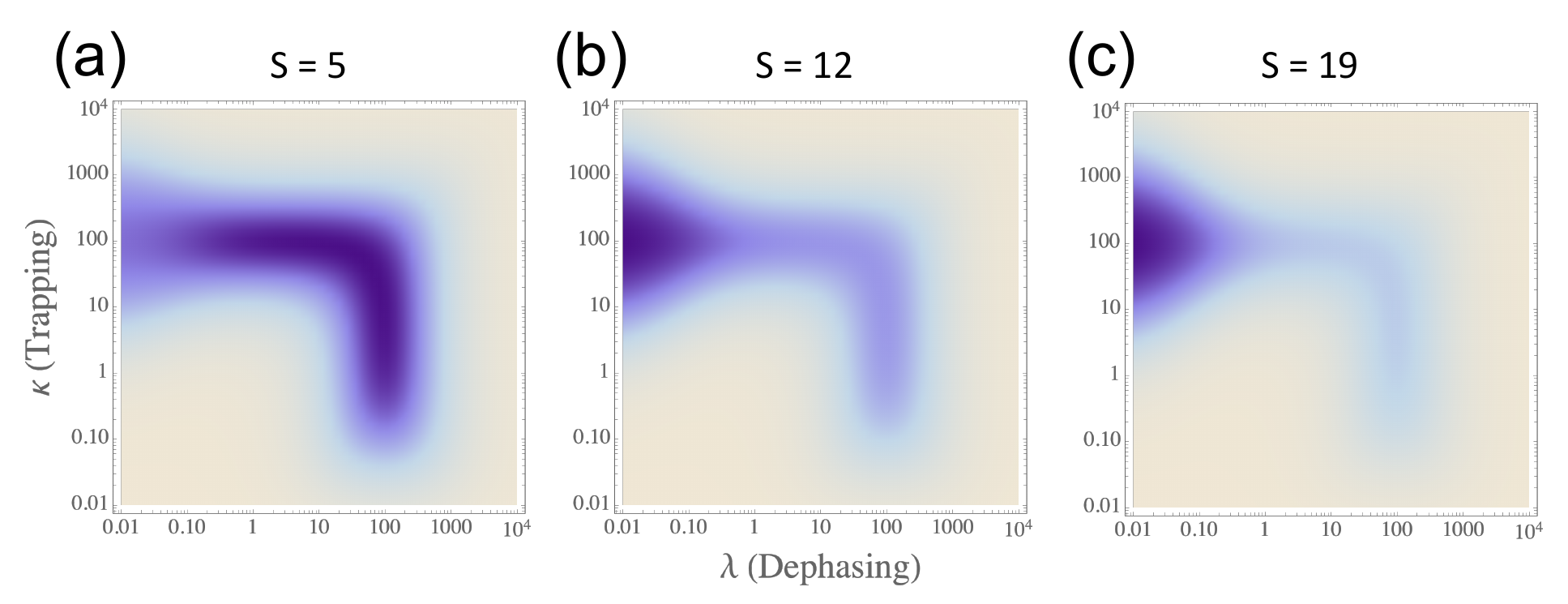} 
\label{fig4}
\caption{Efficiency $\eta$ as a function of $\lambda$ and $\kappa$ for transport on a fully connected network with $N=20$, $\Delta = 100$, $\Gamma = 0.01$, $J=1$ for (a) $S=5$, with optimum $\eta = 0.33$ at $\lambda = 43$ and $\kappa = 58$, (b) $S=12$, with optimum $\eta = 0.57$ at $\lambda = 0$ and $\kappa = 100$ and (c) $S=19$, with optimum $\eta = 0.9$ at $\lambda = 0$ and $\kappa = 100$ }
\end{figure*}

As the initial state coherence increases, the region of maximum efficiency moves from the ENAQT regime, with $\lambda >0$, to the coherent regime, with $\lambda = 0$.  In this regime,  the efficiency can be written in a form similar to Eq.~(\ref{etaweakdecay}), where $\mathcal{R}_0$ is replaced by Eq.~(\ref{Rcoh}).  In particular, one can show that the maximum efficiency for $S=N-1$ occurs for $\kappa = \sqrt{2 (N-1)J^2 + \Delta^2}$, and
\begin{equation}
\eta_{\text{max}} \approx \frac{1}{1+2 \Gamma/\mathcal{R}_{\text{max}}} \ \ \mbox{for} \ \ \lambda=0, S=N-1,
\end{equation}
where $\mathcal{R}_{\text{max}}$ is given by Eq.~(\ref{Rmax}).  This form for $\eta_{\text{max}}$ agrees with the optimal value for Fig.~4(c).  Thus, for sufficiently slow decay, this coherent regime reproduces the behavior of the continuous-time Grover search algorithm.  Note that this coherent regime is distinct from the ENAQT regime, and the order of limits of zero dephasing and slow decay matters for both the efficiency and the transfer rate.

\section{Optimal conditions for ENAQT}
Finally, we return to the expression for the efficiency given by Eq.~(\ref{etaweakdecay}), valid in the slow decay limit, and consider the traditional initial condition with $S=1$.  Using this, and the expressions for $\beta_1$ and $\beta_2$ in Eqs.~(\ref{beta1nodecay})-\ref{beta2nodecay}), we can obtain optimal conditions for the trapping rate $\kappa$ and the dephasing rate $\lambda$.   For $\kappa$, we calculate $\partial \mathcal{R}_0 / \partial \kappa$ and set it equal to zero, which yields the equation:
\begin{equation}
\frac{N J^2}{\kappa^2} = 1 - \frac{\Delta^2}{(\kappa+\lambda)^2}.
\label{dRdkappa}
\end{equation}
This can be solved for small $\lambda$ to find
\begin{equation}
\kappa \to \sqrt{N J^2 + \Delta^2} \ \ \mbox{for} \ \ \lambda \to 0.
\end{equation}
This provides a good approximation for the enhancement of the efficiency in the upper left region of Fig.~2(a).  Similarly, for $\lambda$, we calculate $\partial \mathcal{R}_0 / \partial \lambda$ and set it equal to zero, which yields
\begin{equation}
\frac{2 (N-2) J^2}{\lambda^2} = 1 - \frac{\Delta^2}{(\kappa+\lambda)^2}.
\label{dRdlambda}
\end{equation}
For small $\kappa$, we find
\begin{equation}
\lambda \to \sqrt{ 2(N-2) J^2 + \Delta^2} \ \ \mbox{for} \ \ \kappa \to 0.
\end{equation}
This provides a good approximation for the enhancement of the efficiency in lower right region of Fig.~2(a).  

To find the global optimum, we solve Eqs. (\ref{dRdkappa}) and (\ref{dRdlambda}) simultaneously for $\lambda$ and $\kappa$.  Comparison shows that 
\begin{equation}
\lambda_{\text{opt}} = C \kappa_{\text{opt}},
\end{equation}
where
\begin{equation}
C = \sqrt{ \frac{2(N-2)}{N}}.
\end{equation}
Using this, we can solve the resulting equation for $\kappa$ to obtain
\begin{equation}
\kappa_{\text{opt}} = \sqrt{N J^2 + (1+C)^{-2} \Delta^2 },
\end{equation}
which yields
\begin{equation}
\mathcal{R}_{\text{opt}} = \frac{J^2}{\sqrt{(1+C)^2 N J^2 + \Delta^2}}.
\label{optRenaqt}
\end{equation}
These are the optimal conditions for ENAQT on the FCN.  These expressions for $\kappa_{\text{opt}}$ and $\lambda_{\text{opt}}$ provide excellent approximations to the optimal values found in Figs.~2 and 3 (and more accurate expressions could be developed as a Taylor series in $\Gamma$).  

Furthermore, these optimal conditions display the convergence of time scales discussed in the Introduction, where the dephasing and trapping rates are of the same magnitude (here with coefficient given approximately by $C \approx \sqrt{2}$) and proportional to the energy gap of the system.  For example, when $\Delta \sim \sqrt{N} J$ we have $\kappa \sim \lambda \sim J \sqrt{N}$ and $\mathcal{R}_{\text{opt}} \sim J / \sqrt{N}$.  Note, however, that this optimal rate for the ENAQT regime is quite distinct from the coherent regime discussed in the previous section.  Indeed, setting $S=N-1$ and $\Delta =0$ in Eq.~(\ref{Rmax}) yields $\mathcal{R}_{\text{max}} \approx J \sqrt{(N-1)/2}$.  

\section{Conclusion}

In this paper we have examined the transport of energy on the fully connected network.  The analytical solution covers the whole range of physical parameters, and optimal conditions for ENAQT on this network reveal a convergence of relevant timescales.  We have further shown that initial coherence can enhance the efficiency and rate of transport.  However, the regimes for ENAQT are distinct from the coherence regime, the latter corresponding to continuous-time Grover search.  Here we place our results into a broader context.

First, the transport dynamics, under optimal conditions for ENAQT, exhibits highly damped oscillations, decaying with rate $(\lambda+2\Gamma+\kappa)$ which is comparable to the characteristic frequencies of $\Delta$ and $J$.  Thus, coherent oscillations are not associated with the optimal transfer rate $\mathcal{R}_{\text{opt}}$.  This agrees with the analysis of FMO \cite{rebentrost2009role,hoyer2010limits}.  It is somewhat surprising to see that this remains true in the coherence regime, for which the maximal transfer rate $\mathcal{R}_{\text{max}}$ involves nearly critically damped oscillations.  Thus, the quantum speedup seen in Grover search does not require coherent oscillations.  In fact, using measurement-induced critical damping to improve quantum search has been proposed before \cite{mizel2009critically}.  A related effect for the hitting time on a hypercube was also studied \cite{kendon2003decoherence}.  

Second, the effect of environment on a Grover search algorithm subject to errors was studied in \cite{novo2018environment}.  They found that, for errors similar to disordered energies, the probability of success could be improved by coupling to a thermal environment, although the characteristic $\sqrt{N}$ speedup was not recovered.  It is tempting to compare this to our identification of distinct ENAQT and coherence regimes, for which only the latter achieves quantum speedup.  However, their model of the environment in \cite{novo2018environment} involves energetic relaxation as opposed to simple dephasing.  Thus, there does not seem to be a simple relationship with the model considered here.

Finally, the fact that quantum speedup requires initial coherence and zero dephasing reinforces the arguments made in \cite{hoyer2010limits}, namely that photosynthetic light-harvesting complexes are more likely to be optimized for efficiency and robustness as opposed to achieving a quantum speedup.  Whether there are systems, natural or artificial, in which ENAQT produces a quantum speedup remains an interesting outstanding problem.  

\acknowledgments

The authors thank Williams College for support.  SA thanks D. Aalberts for comments on an earlier draft of this work, and FWS thanks A. Aspuru-Guzik for discussions on and guidance towards ENAQT.

\appendix

\section{Reduced equations of motion}

For completeness, we review the dimensional reduction of the FCN presented in \cite{caruso2009highly}.  We begin by expanding Eq.~(\ref{dmateq}) in terms of the density matrix elements $\rho_{jk}$.  We start by observing that
\begin{align}
(H \rho)_{jk}=&(\varepsilon_j-J)\rho_{jk}+J\sum_{\ell=1}^N\rho_{\ell k}-i(\Gamma+\delta_{j,N}\kappa)\rho_{jk}, \nonumber \\
(\rho H^\dagger)_{jk}=&(\varepsilon_k-J)\rho_{jk}+J\sum_{\ell=1}^N\rho_{j \ell}+i(\Gamma+\delta_{k,N}\kappa)\rho_{jk}. \nonumber \\
\end{align}
We thus find that the total equation of motion for $\rho$ can be written as
\begin{equation}
\dot{\rho}_{jk} =-i(\varepsilon_j-\varepsilon_k)\rho_{jk} - iJ\sum_{\ell=1}^N (\rho_{\ell k}- \rho_{j \ell}) -\Lambda_{j,k}\rho_{jk}, \label{eq:rhodot}
\end{equation}
with $\Lambda_{j,k}$ representing the dephasing, decay, and trapping terms given by
\begin{equation}
\Lambda_{j,k}=(1-\delta_{j,k})\lambda + 2\Gamma+(\delta_{j,N}+\delta_{k,N})\kappa. \label{eq:bigLdef}
\end{equation}

To reduce the number of equations, we set $j=k=N$ in Eq.~(\ref{eq:rhodot}) to find
\begin{equation}
\dot{\rho}_{NN} = - iJ \sum_{\ell=1}^N (\rho_{ \ell N}-\rho_{N \ell} )-(2 \Gamma + 2 \kappa) \rho_{NN}.
\end{equation} 
Here we see transport is mediated by the sum of coherences with state $N$.  This motivates the introduction of the auxiliary variable
\begin{equation}
A_k = \sum_{j=1}^N \rho_{jk}, 
\end{equation}
in terms of which 
\begin{equation}
\dot{\rho}_{NN} = - i J (A_N-A^\ast_N) -(2 \Gamma + 2 \kappa) \rho_{NN}.
\end{equation} 
We further define their sum
\begin{equation}
S=\sum_{k=1}^N A_k =\sum_{k=1}^N A^\ast_k=\sum_{j,k=1}^N \rho_{jk} \label{eq:dephsdef},
\end{equation}
and the trace of the density matrix
\begin{equation} 
T = \text{Tr}(\rho) = \sum_{k=1}^N \rho_{kk}.
\end{equation}

We now proceed to find the equations for these variables, starting with $A_k$.  We calculate
\begin{align}
\dot{A}_k=\sum_{j=1}^N \dot{\rho}_{jk} =& -i\sum_{j=1}^N(\varepsilon_j-\varepsilon_k)\rho_{jk} -iJ(NA_k-S) \nonumber \\
 & -\sum_{j=1}^N\Lambda_{j,k}\rho_{jk}\label{eq:adot}.
\end{align}
The last term can be expanded using Eq.~(\ref{eq:bigLdef})
\begin{equation}
\sum_{j=1}^N \Lambda_{j,k} \rho_{jk} = ( \lambda + 2 \Gamma) A_k - \lambda \rho_{kk} + \kappa \rho_{Nk} + \kappa \delta_{k,N} A_k.
\end{equation}
Using this result and $\varepsilon_j = \varepsilon_N \delta_{j,N}$ in Eq.~(\ref{eq:adot}), we find
\begin{align}
\dot{A}_k =& -(\kappa+i\varepsilon_N)\rho_{Nk} - (\kappa-i\varepsilon_N) \delta_{N,k} A_k \nonumber \\
 & - i J(N A_k-S) - (\lambda + 2 \Gamma) A_k + \lambda\rho_{kk}. \label{eq:generaladot}
\end{align}
While this depends on the individual coherences $\rho_{Nk}$, recall that $\dot{\rho}_{NN}$ depends on $A_N$.  Thus, when $k=N$ in Eq.~(\ref{eq:generaladot}), only $\rho_{NN}$, $A_N$, and $S$ appear on the right-hand-side.

We proceed to find the equation for $S$, using our results for $\dot{A}_k$:
\begin{align}
\dot{S} =\sum_{k=1}^N \dot{A}_{k} =& -(\kappa +i \varepsilon_N) A^\ast_N - (\kappa - i \varepsilon_N) A_N \nonumber \\
& - (\lambda + 2 \Gamma) S + \lambda T. 
\label{eq:sdot}
\end{align}
Finally, we calculate
\begin{equation}
\dot{T} =- 2 \Gamma T - 2 \kappa \rho_{NN}.
\end{equation}
This completes the system of equations.  To simplify, we  set $A_N = X + i Y$ in Eq.~(\ref{eq:generaladot}) to find
\begin{align}
\dot{X} &= \text{Re}(\dot{A}_N) \nonumber \\
&= - (\kappa - \lambda) \rho_{NN} - (\lambda + 2 \Gamma + \kappa) X + (JN-\varepsilon_N) Y \nonumber \\
\end{align}
and
\begin{align}
\dot{Y} &= \text{Im}(\dot{A}_N) \nonumber \\
&= - \varepsilon_N \rho_{NN} - (JN-\varepsilon_N) X - (\lambda + 2 \Gamma + \kappa) Y + J S.  \nonumber \\
\end{align}

Altogether we have the closed system of equations
\begin{align}
\dot{\rho}_{NN}=&-2(\Gamma+\kappa)\rho_{NN}+2JY,\nonumber \\
\dot{X}=&-(\kappa-\lambda)\rho_{NN}-(\lambda+2\Gamma+\kappa)X + (JN-\varepsilon_N) Y,\nonumber \\
\dot{Y}=&-\varepsilon_N \rho_{NN} -(JN-\varepsilon_N) X-(\lambda+2\Gamma+\kappa)Y+JS,\nonumber \\
\dot{S}=&-2\kappa X-2\varepsilon_N Y-(\lambda+2\Gamma)S+\lambda T, \nonumber \\
\dot{T}=&-2\kappa \rho_{NN}-2\Gamma T.
\end{align}

\bibliography{enaqt}

\end{document}